\documentclass[12pt,showpacs,double]{revtex4}
\usepackage{amsmath}
\usepackage{graphics}
\usepackage{graphicx}

\input{tcilatex}

\begin{document}

\title{Abrupt barrier contribution to the electron spin splitting in
asymmetric coupled double quantum wells.}
\author{A. Hern\'{a}ndez-Cabrera}
\email{ajhernan@ull.edu.es}
\author{P. Aceituno}
\email{paceitun@ull.edu.es}
\affiliation{Dpto. F{\'{\i}}sica B\'{a}sica, Universidad de La Laguna, La Laguna,
38206-Tenerife, Spain, and Instituto Universitario de Estudios Avanzados
(IUdEA) en F\'{\i}sica At\'{o}mica, Molecular y Fot\'{o}nica, Universidad de
La Laguna, La Laguna, 38206 Tenerife, Spain}
\date{\today }

\begin{abstract}
We have studied the behavior of the electronic energy spin-splitting of $%
InGaAs-InAlAs$ based double quantum wells (narrow gap structures) under
in-plane magnetic and transverse electric fields. We have developed an
improved $8\times 8$ version of the Transfer Matrix Approach that consider
contributions from abrupt interfaces and external fields when tunneling
through central barrier exists. We have included the Land\'{e} $g$-factor
dependence on the external applied field. Also, we have calculated electron
density of states and photoluminescence excitation. Variations of the
electron spin-splitting energy lead to marked peculiarities in the density
of states. Because the density of states is directly related to
photoluminescence excitation, these peculiarities are observable by this
technique.
\end{abstract}

\pacs{72.25.-b, 73.21.-b }
\keywords{Double quantun well; spin; abrupt interface; pholuminescence
excitation.}
\maketitle

\newpage

\section*{1. Introduction}

In the last decade a great interest has arisen for the so-called spintronic,
or spin-based electronics. The reason is that the spintronic, due to its low
power consumption, promises to be a good alternative to the traditional
electronics, based on the charge transport \cite{1}. On the other hand, the
spin transport is not very dissipative, with low energy losses over long
distances, although the relaxation of the spin polarization may exist.
Spintronics is a catchall term that refers to the potential use of the spin
rather than the charge in electronic devices. For it to be useful in
practice, spin up and spin down electronic states (or hole states) of any
material must be separated in energy. Also, the material should be
electrically polarized as in conventional electronics, which means that
carriers, both negative (electrons) and positive (holes), must be able to
conduct.

A key point is the choice of the material. For example, narrow gap
semiconductors with strong spin-orbit coupling. $GaAs$-based materials seem
to be the suitable candidates\cite{2} due to the long life of the magnetic
spin state of photoexcited electrons, which behaves coherently. It is also
important that the materials have a good lattice matching to avoid defects
accumulation at the interface and internal strains, which would worsen the
transport of the polarized spin. There exist techniques to relax this
strain, as to place step graded buffers with a progressive variation of the $%
In$ concentration between the substrate and the active device. These layers
mainly absorb strain and defects.

Another important point is a large Land\'{e} factor for having a significant
splitting of spin states by applying small magnetic fields. In general, $%
In_{x}Ga_{1-x}As-In_{y}Al_{1-y}As$ structure seems to be one of the most
appropriate for spintronic purposes. This heterostructure offers the
possibility of manipulate the gap width, the Land\'{e} factor and the
interface contributions by varying $x$ and $y$ concentrations. Besides, it
presents a remarkable spin-splitting energy when a weak magnetic fields is
applied. This peculiarity makes this material suitable for high temperature
spin-valves devices.

In this work we will focus in the spin-splitting changes in asymmetric
coupled double quantum wells ACQW, caused by abrupt interfaces, when an
in-plane magnetic field is applied. For this purpose we will base on an
extended version of the $8\times 8$ Kane formalism with nonsymmetric
boundary conditions to calculate the band structure and dispersion laws\cite%
{3}. Although electronic dispersion laws (and the corresponding spin-orbit
splitting) in quantum heterostructures have been widely studied in the last
decades\cite{4}, many question remain open. One of them is the influence of
heterojunctions on the electron spin tunneling and thus, their contribution
to the polarization of this spin. In theory, effects from compositional
parameters or interface contributions to the spin splitting should be
analyzed through the density of states. And the modifications of the density
of states can be directly observed using photoluminescence excitation
technique (PLE)\cite{5}.

In bulk materials spin-orbit interaction is caused either by a soft
potential \cite{6} and by cubic \cite{7} and linear \cite{8} spin-dependent
contributions to the effective Hamiltonian. However, in the two-dimensional
(2D) case, we can reduce the cubic contribution to a linear one after the
squared momentum substitution by its quantized value due to confinement \cite%
{9}. Moreover, it is necessary to consider the additional spin-orbit
splitting caused by the interaction with abrupt heterojunction potentials
(see Ref. \cite{1,3,10,11}). This contribution is absolutely different to
the contributions mentioned above.

The effect of an\ in-plane magnetic field on the energy spectrum in
nonsymmetric heterostructures results in the Pauli contribution to the
electron Hamiltonian. Several peculiarities for transport phenomena in
heterostructures have been also discussed \cite{12,13,14,15}. However, these
discussions only consider the mixing between Pauli contribution and
effective 2D spin-orbit interaction. As mentioned above, we cannot forget
the contribution of the abrupt barriers in asymmetric structures (e.g. ACQW)
to the spin splitting, as we will emphasize in this work. Besides, we have
included possible changes of the Land\'{e} factor. Although the Land\'{e} $g$%
-factor depends on the applied fields \cite{16,17}, this dependence is
negligible for in-plane magnetic fields and low electron density. The model
also applies to narrow-gap heterostructures whenever the slow potential \
generated by doping or external transverse electric fields can be described
self-consistently.

\section*{2. Eigenstate problem}

We will center our attention in the Hamiltonian describing the electronic
behavior in the conduction band, considering that any possible strain is
already included in the structure through the gap, the conduction and the
valence well potentials\cite{18}. We will include effects of electric and
magnetic fields as well as the interfaces contribution.

Based on the assumption that the gap energy $\varepsilon _{g}$ is smaller
than the energy distance between the valence band ($v$-band) and the
spin-split band extrema, the electronic states in these narrow-gap
heterostructures can be described by the three-band Kane matrix Hamiltonian 
\begin{equation}
\hat{\varepsilon}(z)+(\hat{\mathbf{v}}\cdot \hat{\mathbf{P}}),~~~~~~\hat{%
\mathbf{P}}=\hat{\mathbf{p}}-\frac{e}{c}\mathbf{A},  \label{1}
\end{equation}%
where the generalized kinetic momentum, $\hat{\mathbf{P}}$, contains the
vector potential $\mathbf{A}=(Hz,0,0)$, $\mathbf{H}\Vert OY$ is an in-plane
magnetic field and $\hat{\mathbf{p}}=\left( \mathbf{p},\hat{p}_{z}\right) $
is written in the $\mathbf{p},z$-representation through the 2D momentum $%
\mathbf{p}$. We have also introduced the diagonal energy matrix $\hat{%
\varepsilon}(z)$ whose elements fix the positions of the band extrema and
the interband velocity matrix $\hat{\mathbf{v}}$.

From now on it is necessary to introduce a new index $\mu =w,b$\ to denote
narrow-gap regions (wells) and wide-gap regions (central barrier and lateral
sides), respectively. In the parabolic approximation we can write the Schr%
\"{o}dinger equation for ACQW in the form\cite{3}: 
\begin{equation}
\left( \varepsilon _{p}^{\mu }+\frac{\hat{p}_{z}^{2}}{2m_{\mu }}+\varepsilon
_{c}^{\mu }(z)+\widehat{W}^{\mu }(z)\right) \Psi ^{\mu }(\mathbf{p,}z\mathbf{%
)}=E\Psi ^{\mu }(\mathbf{p,}z\mathbf{)},  \label{2}
\end{equation}%
where the isotropic kinetic energy is given by 
\begin{equation}
\varepsilon _{p}^{\mu }=\frac{p_{x}^{2}+p_{y}^{2}}{2m_{\mu }},  \label{3}
\end{equation}%
which includes the effective mass $m_{\mu }$. Parabolic approximation is
justified because energy values under consideration are smaller than the gap
energy $\varepsilon _{g}$ in the narrow region. The $\varepsilon _{c}^{\mu
}(z)$ energy is $\varepsilon _{c}^{w}(z)=U(z)$ in the wells, and $%
\varepsilon _{c}^{b}(z)=\Delta E_{c}+U(z)$ in the barriers, where $\Delta
E_{c}$ is the band offset for $c$ conduction band. Whenever energy values
{}{}are less than $\Delta E_{c}$ underbarrier penetration (and tunneling) is
permitted and described by the boundary conditions.\ The potential $U(z)$,
for an uniform transverse electric field, is $U(z)\simeq eF_{\bot }z$. Band
diagram for ACQW is shown in Fig. 1.

The magnetic energy $\widehat{W}^{\mu }(z)$, for not very strong magnetic
fields, is described by%
\begin{equation}
\widehat{W}^{\mu }(z)=-V^{\mu }(z)\left[ \mathbf{\hat{\sigma}}\times \mathbf{%
p}\right] _{z}+\frac{g^{\mu }(z)}{2}\mu _{B}H\hat{\sigma}_{y},  \label{4}
\end{equation}%
where $\mu _{B}\equiv |e|\hbar /(m_{e}c)$ is the Bohr magneton and $\mathbf{%
\hat{\sigma}}$ is the Pauli matrix. Finally, the characteristic spin
velocity $V^{\mu }(z)$, and the effective Land\'{e} factor $g^{\mu }(z)$\
are 
\begin{equation}
V^{\mu }(z)=\frac{\hbar }{4m_{\mu }}\frac{d\varepsilon _{c}^{\mu }(z)/dz}{%
\varepsilon _{g}},\ \ g^{\mu }(z)=\frac{m_{e}}{2m_{\mu }}\left[ 1+z\frac{%
d\varepsilon _{c}^{\mu }(z)/dz}{\varepsilon _{g}}\right] .  \label{5}
\end{equation}

The potential of interfaces determines a part of the spin dependent
contributions through the parameter $\chi $. \ Actually, $\chi $ takes into
account the spin-orbit coupling due to the abrupt potential of the
heterojunction at each interface, as\cite{3,11} 
\begin{equation}
\chi =\frac{2}{\hbar }\int_{-\delta }^{\delta }m_{\mu }V^{\mu }(z)\ dz,
\label{6}
\end{equation}%
where the integral is taken over the width of the abrupt interface $2\delta $%
. Contribution of $\chi $ to the energy dispersion relations will appear
through the third kind boundary conditions after a first integration of the
Schr\"{o}dinger equation (2). We will detail this point in Appendix A.

In the present case, where $U(z)$ is linear with $z$, \ we get $d\varepsilon
_{c}^{\mu }(z)/dz=dU(z)/dz=eF_{\bot }$, which does not depend on $z$. Thus,\
the external applied electric field $F_{\bot }$ shapes the spin velocity $%
V^{\mu }(z)=\overline{\mathrm{v}}^{\mu }$ and the $g$-factor, $g^{\mu
}(z)=g^{\mu }$. \ We can take the characteristic spin velocity for each
layer as 
\begin{equation}
\overline{\mathrm{v}}^{\mu }=\frac{eF_{\bot }\hbar }{4m_{\mu }\varepsilon
_{g}},  \label{7}
\end{equation}%
and the abrupt interface parameter as 
\begin{equation}
\chi =\frac{2eF_{\bot }\delta +\Delta E_{c}}{2\varepsilon _{g}}\approx \frac{%
\Delta E_{c}}{2\varepsilon _{g}}  \label{8}
\end{equation}%
Lastly, we introduce the Pauli splitting energy $w_{H}$, caused by the
magnetic field, as $w_{H}^{\mu }=(g^{\mu }/2)\mu _{B}H$. Thus, Eq. (4)
becomes 
\begin{equation}
\widehat{W}^{\mu }=\bar{\mathrm{v}}^{\mu }\left[ \mathbf{\hat{\sigma}}\times 
\mathbf{p}\right] _{z}+w_{H}^{\mu }\hat{\sigma}_{y},  \label{9}
\end{equation}%
which has lost the $z$ dependence.

Because $\widehat{W}^{\mu }$ and $\varepsilon _{p}^{\mu }$ do not depend on $%
z$, we can factorize fundamental solutions of Eq. (2), $\Psi ^{\mu }(\mathbf{%
p,}z\mathbf{)}$, as products of $\ \psi ^{\mu \sigma }(\mathbf{p)}$ and $%
\varphi ^{\mu \sigma }(z)$ functions, with $\sigma =\pm 1$. The $\sigma $
value refers to the two possible spin orientations. For an ACQW under a
transverse electric field $F_{\bot }$, the $\mathbf{p}$-dependent spinors $%
\psi ^{\mu \sigma }(\mathbf{p)}$ can be obtained from%
\begin{equation}
\left( \varepsilon _{p}^{\mu }+\widehat{W}^{\mu }\right) \psi ^{\mu \sigma }(%
\mathbf{p)}=\varepsilon _{\sigma \mathbf{p}}^{\mu }\psi ^{\mu \sigma }(%
\mathbf{p)},  \label{10}
\end{equation}%
in the form 
\begin{eqnarray}
\psi ^{\mu +}(\mathbf{p)} &=&\frac{1}{\sqrt{2}}\left\vert 
\begin{array}{l}
~~~~~~~~~~~1 \\ 
(\overline{\mathrm{v}}^{\mu }p_{+}+w_{H}^{\mu })/i\mathrm{w}_{\mathbf{p}%
}^{\mu }%
\end{array}%
\right\vert ~~,  \notag \\
\psi ^{\mu -}(\mathbf{p)} &=&\frac{1}{\sqrt{2}}\left\vert 
\begin{array}{l}
(\overline{\mathrm{v}}^{\mu }p_{-}+w_{H}^{\mu })/i\mathrm{w}_{\mathbf{p}%
}^{\mu } \\ 
~~~~~~~~~~~1%
\end{array}%
\right\vert ~~,  \label{11}
\end{eqnarray}%
where%
\begin{equation}
p_{+}=p_{x}+ip_{y},\ \text{ }p_{-}=p_{x}-ip_{y}\text{ and }\mathrm{w}_{%
\mathbf{p}}^{\mu }=\sqrt{(\bar{\mathrm{v}}^{\mu }p_{x}+w_{H}^{\mu })^{2}+(%
\overline{\mathrm{v}}^{\mu }p_{y})^{2}}\text{ }  \label{12}
\end{equation}%
and the energy quasi-paraboloids are 
\begin{equation}
\varepsilon _{\mathbf{p}\sigma }^{\mu }=\varepsilon _{p}^{\mu }+\sigma |%
\mathrm{w}_{\mathbf{p}}^{\mu }|,~~~~  \label{13}
\end{equation}

Now, $z$-dependent \ functions $\varphi ^{\mu \sigma }(z)$ are obtained from
the second order differential equation 
\begin{equation}
\left( \frac{\hat{p}_{z}^{2}}{2m_{\mu }}+eF_{\bot }z\right) \varphi ^{\mu
\sigma }(z)=\left( E-\varepsilon _{\mathbf{p}\sigma }^{\mu }\right) \varphi
^{\mu \sigma }(z).  \label{14}
\end{equation}%
For an ACQW under a transversal electric field $F_{\bot }$, eigenstate
functions of Eq. (14) are the known linear combination of the Airy $Ai$- and 
$Bi$-functions. Thus, the general solution of \ Eq. (2) can be written as 
\begin{eqnarray}
\Psi ^{\mu }(\mathbf{p,}z) &=&\psi ^{\mu +}(\mathbf{p})\left[ a_{\mu
+}A_{i}^{\mu +}(z)+b_{\mu +}B_{i}^{\mu +}(z)\right] +  \notag \\
&&+\psi ^{\mu -}(\mathbf{p})\left[ a_{\mu -}A_{i}^{\mu -}(z)+b_{\mu
-}B_{i}^{\mu -}(z)\right] ,  \label{15}
\end{eqnarray}%
where $a_{\mu \sigma },\ b_{\mu \sigma }$ are four constants by region to be
solved through the interface conditions of continuity of the wave functions
and current. Because there are two wells, and three barriers (Fig. 1), there
are four interfaces and four boundary conditions by interface. \ 

Next, to simplify the calculation of the wave functions and the dispersion
relations of the electronic levels through the boundary conditions\cite{3,10}%
, we create two auxiliary parameters: a length $l_{\bot }^{\mu }$ and an
energy $\varepsilon _{\bot }^{\mu }$ 
\begin{equation}
l_{\bot }^{\mu }=\left( \frac{\hbar ^{2}}{2m_{\mu }eF_{\bot }}\right)
^{1/3},\ \ \varepsilon _{\bot }^{\mu }=\frac{\hbar ^{2}}{2m_{\mu }\left(
l_{\bot }^{\mu }\right) ^{2}},  \label{16}
\end{equation}%
and a set of momentum dependent functions%
\begin{eqnarray}
\rho _{\mathbf{p-}}^{\mu } &=&\frac{p_{-}\bar{\mathrm{v}}^{\mu }+w_{H}^{\mu }%
}{i\mathrm{w}_{\mathbf{p}}^{\mu }},\text{ with }\rho _{\mathbf{p-}}^{b}=\rho
_{\mathbf{p-}}^{w}=\rho _{\mathbf{p-}}  \notag \\
\rho _{\mathbf{p+}}^{\mu } &=&\frac{p_{+}\bar{\mathrm{v}}^{\mu }+w_{H}^{\mu }%
}{i\mathrm{w}_{\mathbf{p}}^{\mu }},\text{ with }\rho _{\mathbf{p+}}^{b}=\rho
_{\mathbf{p+}}^{w}=\rho _{\mathbf{p+}}  \notag \\
f_{1\mathbf{p}} &=&i\chi \frac{p_{-}}{\hbar }\rho _{\mathbf{p+}},  \notag \\
f_{2\mathbf{p}} &=&i\chi \frac{p_{-}}{\hbar },  \notag \\
f_{3\mathbf{p}} &=&-i\chi \frac{p_{+}}{\hbar },  \notag \\
f_{4\mathbf{p}} &=&-i\chi \frac{p+}{\hbar }\rho _{\mathbf{p-}}.  \label{17}
\end{eqnarray}%
Next, using preliminary quasi parabolic dispersion relations $\varepsilon _{%
\mathbf{p}\sigma }^{\mu }$, we construct the Airy function arguments%
\begin{equation}
\xi _{\mathbf{p}\sigma }^{\mu }=\frac{z}{l_{\bot }^{\mu }}+\frac{\varepsilon
_{\mathbf{p}\sigma }^{\mu }-E+\delta _{b}^{\mu }\Delta E_{c}}{\varepsilon
_{\bot }^{\mu }},  \label{19}
\end{equation}%
where $\delta _{\mu }^{b}$ acts as a Kronecker function: $\delta _{\mu
}^{b}=1$ when $\mu =b$, and $\delta _{\mu }^{b}=0$ when $\mu =w$.

We use the transfer matrix method to obtain wave functions and energy
dispersion relations of the electronic levels. In the present case, we have
improved the standard method by using $4\times 4$ matrices at each
interface, whose elements are (see Appendix A):%
\begin{eqnarray}
\mathcal{M}_{11}^{\mu }(z,E,\mathbf{p}) &=&Ai\left( \xi _{\mathbf{p+}}^{\mu
}\right) ,  \notag \\
\mathcal{M}_{12}^{\mu }(z,E,\mathbf{p}) &=&Bi(\xi _{\mathbf{p+}}^{\mu }), 
\notag \\
\mathcal{M}_{13}^{\mu }(z,E,\mathbf{p}) &=&\rho _{\mathbf{p-}}Ai(\xi _{%
\mathbf{p-}}^{\mu }),  \notag \\
\mathcal{M}_{14}^{\mu }(z,E,\mathbf{p}) &=&\rho _{\mathbf{p-}}Bi(\xi _{%
\mathbf{p-}}^{\mu }),  \notag \\
\mathcal{M}_{21}^{\mu }(z,E,\mathbf{p}) &=&\rho _{\mathbf{p+}}Ai(\xi _{%
\mathbf{p+}}^{\mu }),  \notag \\
\mathcal{M}_{22}^{\mu }(z,E,\mathbf{p}) &=&\rho _{\mathbf{p+}}Bi(\xi _{%
\mathbf{p+}}^{\mu }),  \notag \\
\mathcal{M}_{23}^{\mu }(z,E,\mathbf{p}) &=&Ai(\xi _{\mathbf{p-}}^{\mu }), 
\notag \\
\mathcal{M}_{24}^{\mu }(z,E,\mathbf{p}) &=&Bi(\xi _{\mathbf{p-}}^{\mu }), 
\notag \\
\mathcal{M}_{31}^{\mu }(z,E,\mathbf{p}) &=&\frac{m_{e}}{m_{\mu }}Ai^{\prime
}(\xi _{\mathbf{p+}}^{\mu })+\delta _{\mu }^{w}f_{1\mathbf{p}}Ai\left( \xi _{%
\mathbf{p+}}^{\mu }\right) ,  \notag \\
\mathcal{M}_{32}^{\mu }(z,E,\mathbf{p}) &=&\frac{m_{e}}{m_{\mu }}Bi^{\prime
}(\xi _{\mathbf{p+}}^{\mu })+\delta _{\mu }^{w}f_{1\mathbf{p}}Bi\left( \xi _{%
\mathbf{p+}}^{\mu }\right) ,  \notag \\
\mathcal{M}_{33}^{\mu }(z,E,\mathbf{p}) &=&\frac{m_{e}}{m_{\mu }}\rho _{%
\mathbf{p-}}Ai^{\prime }(\xi _{\mathbf{p-}}^{\mu })+\delta _{\mu }^{w}f_{2%
\mathbf{p}}Ai(\xi _{\mathbf{p-}}^{\mu }),  \notag \\
\mathcal{M}_{34}^{\mu }(z,E,\mathbf{p}) &=&\frac{m_{e}}{m_{\mu }}\rho _{%
\mathbf{p-}}Bi^{\prime }(\xi _{\mathbf{p-}}^{\mu })+\delta _{\mu }^{w}f_{2%
\mathbf{p}}Bi(\xi _{\mathbf{p-}}^{\mu }),  \notag \\
\mathcal{M}_{41}^{\mu }(z,E,\mathbf{p}) &=&\frac{m_{e}}{m_{\mu }}\rho _{%
\mathbf{p+}}Ai^{\prime }(\xi _{\mathbf{p+}}^{\mu })+\delta _{\mu }^{w}f_{3%
\mathbf{p}}Ai\left( \xi _{\mathbf{p+}}^{\mu }\right) ,  \notag \\
\mathcal{M}_{42}^{\mu }(z,E,\mathbf{p}) &=&\frac{m_{e}}{m_{\mu }}\rho _{%
\mathbf{p+}}Bi^{\prime }(\xi _{\mathbf{p+}}^{\mu })+\delta _{\mu }^{w}f_{3%
\mathbf{p}}Bi\left( \xi _{\mathbf{p+}}^{\mu }\right) ,  \notag \\
\mathcal{M}_{43}^{\mu }(z,E,\mathbf{p}) &=&\frac{m_{e}}{m_{\mu }}Ai^{\prime
}(\xi _{\mathbf{p-}}^{\mu })+\delta _{\mu }^{w}f_{4\mathbf{p}}Ai\left( \xi _{%
\mathbf{p-}}^{\mu }\right) ,  \notag \\
\mathcal{M}_{44}^{\mu }(z,E,\mathbf{p}) &=&\frac{m_{e}}{m_{\mu }}Bi^{\prime
}(\xi _{\mathbf{p-}}^{\mu })+\delta _{\mu }^{w}f_{4\mathbf{p}}Bi\left( \xi _{%
\mathbf{p-}}^{\mu }\right) ,  \label{20}
\end{eqnarray}%
where $Ai^{\prime }(\xi _{\sigma \mathbf{p}}^{\mu })$ means $dAi(\xi
_{\sigma \mathbf{p}}^{\mu })/dz=\left( 1/l_{\perp }^{\mu }\right) \ dAi(\xi
_{\sigma \mathbf{p}}^{\mu })/d\xi _{\sigma \mathbf{p}}^{\mu }$, and the same
for $Bi^{\prime }(\xi _{\sigma \mathbf{p}}^{\mu }).$

Now we are ready to generate transfer matrices, $\widetilde{M}^{\mu }(z,E,%
\mathbf{p})$ which elements are $\mathcal{M}_{ij}^{\mu }(z,E,\mathbf{p}).$
Finally, electronic levels for each 2D momentum $\mathbf{p}=\left(
p_{x},p_{y}\right) $ are obtained from $\widetilde{S}_{44}\left( E,\mathbf{p}%
\right) =0$, where

\begin{eqnarray}
\widetilde{S}\left( E,\mathbf{p}\right) &=&\left[ \widetilde{M}^{b}(L_{0},E,%
\mathbf{p})\right] ^{-1}\cdot \widetilde{M}^{w}(L_{0},E,\mathbf{p})\cdot %
\left[ \widetilde{M}^{w}(L_{1},E,\mathbf{p})\right] ^{-1}\cdot  \notag \\
&&\widetilde{M}^{b}(L_{1},E,\mathbf{p})\cdot \left[ \widetilde{M}%
^{b}(L_{2},E,\mathbf{p})\right] ^{-1}\cdot \widetilde{M}^{w}(L_{2},E,\mathbf{%
p})\cdot  \notag \\
&&\left[ \widetilde{M}^{w}(L_{3},E,\mathbf{p})\right] ^{-1}\cdot \left[ 
\widetilde{M}^{b}(L_{3},E,\mathbf{p})\right] .  \label{21}
\end{eqnarray}%
In the above matrix product, $z=L_{i}$ denotes interfaces position in the
growth direction, starting from the left side.

Calculations give us two spin up $E_{k+}(\mathbf{p})$ and two spin down $%
E_{k-}(\mathbf{p})$ paraboloids,$\ $where $k=1,2$ corresponds to the
deepest\ coupled levels of the ACQW. Once obtained coefficients $a_{\mu
\sigma },\ b_{\mu \sigma }$\ (Eq. 15) we normalize wave functions for each
momentum $\mathbf{p}$.

The scheme of Fig. 1 includes the two resonant energy levels and the
respective wave functions for $\mathbf{p=0}$. Although there are four levels
only two are observable in this figure. This is because spin sublevel
splitting is much smaller than electronic level energy distance and
differences between spin down and spin up wave functions are not visible at $%
\mathbf{p=0}$. \ 

Finally, the density of states can be obtained by using the well-known
expression

\begin{equation}
\rho (\varepsilon )=\sum_{k,\sigma }\int \frac{d\mathbf{p}}{(2\pi \hbar )^{2}%
}\delta (\varepsilon -E_{k\sigma }(\mathbf{p})).  \label{22}
\end{equation}%
Peculiarities of $\rho (\varepsilon )$ can be analyzed experimentally
through the photoluminescence excitation (PLE) intensity for the case of
near-edge absorption, $I_{PLE}$, because both quantities are related by\cite%
{11} 
\begin{equation}
I_{PLE}\sim \sum_{\lambda _{c}\lambda _{v}}|\mathbf{e\cdot v}%
_{cv}|^{2}\delta (\varepsilon _{\lambda _{c}}-\varepsilon _{\lambda
_{v}}-\hbar \omega )\sim \rho (\hbar \Delta \omega )  \label{23}
\end{equation}%
for very low temperature, or 
\begin{equation}
I_{PLE}\sim \sum_{\lambda _{c}\lambda _{v}}|\mathbf{e\cdot v}%
_{cv}|^{2}G(\varepsilon _{\lambda _{c}}-\varepsilon _{\lambda _{v}}-\hbar
\omega ),  \label{24}
\end{equation}%
when including electron-phonon scattering. In the above expression, the
Gaussian function is $G(x)=\frac{1}{\gamma \sqrt{2\pi }}\exp \left[ -\left( 
\frac{x-x_{0}}{\gamma \sqrt{2}}\right) ^{2}\right] $. The Gaussian halfwidth 
$\gamma $ is related to the scattering and relaxation processes and, thus,
to the temperature\cite{19}. Expressions (23, 24) are valid provided the
interband velocity $\mathbf{v}_{cv}$ does not depend on in-plane momentum.
Here $\mathbf{e}$\textbf{\ }is the light polarization vector, $\Delta \omega
=\omega -\varepsilon _{g}/\hbar $, and $\varepsilon _{\lambda _{c}}$, $%
\varepsilon _{\lambda _{v}}$ are the conduction and valence band levels,
respectively.

\section*{3. Results}

Let's start this section with numerical results for $%
In_{x}Ga_{1-x}As-In_{y}Al_{1-y}As$-based ACQWs, with $x=0.53$ and $y=0.52$.
We have chosen this particular structure because we have reliable data for
basic parameters\cite{20,21}. We have considered two $InGaAs$ wells of $70$
and $100$ \AA\ wide separated by a $20$ \AA\ $InAlAs$ barrier. We have also
applied an electric field of $30$ kV/cm, which corresponds to a spin
velocity $\overline{\mathrm{v}}^{w}=2.6\times 10^{5}$ cm/s for the $InGaAs$
QWs, and $\overline{\mathrm{v}}^{b}=1.4\times 10^{5}$ cm/s for the $InAlAs$
barriers, with a transition spin velocity region across de abrupt interface.
This electric field is slightly higher than needed to achieve resonance
between the deepest levels of both wells ($28$ kV/cm). To calculate
interface contributions we have used a typical abrupt interface size of $%
\delta \sim 3\mathring{A}$\ for $InGaAs-InAlAs$\cite{22}. We have also
applied in-plane magnetic field of $0.01\ $T, small enough to allow the
anticrossing close to the bottom of energy dispersion relations (zero slope
points).

Figures 2$(a-d)$ show normalized squared wave function for $p_{y}=0$, versus 
$z$ and the dimensionless momentum $p_{x}/p_{0}$, where $p_{0}=m_{w}%
\overline{\mathsf{v}}^{w}$\textrm{.} Upper panels $(a,b)$ correspond to the
first deepest level for the two different spin orientations. As expected for
an electric field beyond the resonance, charge density is mainly located in
the left narrow QW. Consequently, the opposed happens for the second
resonant level as can be seen in the lower panels $(c,d)$ Analyzing wave
functions versus momentum $p_{x}$ and spin orientations by comparing panels $%
(a)$ and $(b)$, a particular behavior occurs. While the charge distribution
coincide for both down and up spins at the zone center ($p_{x}=0$), there is
a tunneling charge transfer between wells for increasing $\left\vert
p_{x}\right\vert $. For spin down case [panel $\left( a\right) $] charge
enhances tunneling from left narrow well to the right wide one. Tunneling
shows opposite behavior for spin up electrons [panel ($b$)] and charge goes
from the right to the left well. Lower panels $(c,d)$ correspond to the
higher resonant level, mainly at the right well. It might seem the behavior
is the opposite to the previous one because now, spin down case [panel $%
\left( c\right) $] shows a tunneling increase from right to left QW as $%
\left\vert p_{x}\right\vert $ increases and, conversely, for spin up
electrons [panel $\left( d\right) $]. However, considering the relative
charge concentration between wells, we can realize there is a similar
behavior for both resonant levels. For spin down electrons \ there is a
charge shift from well with higher concentration to the other well [panels $%
(a)$, $(c)$] and conversely for the spin up electrons [panels $(b)$, $(d)$].
The reason for this charge shift lies in the magnetic energy term $w_{%
\mathbf{p}}^{\mu }$, which induces a breaking of the \ $p_{x}$ momentum
symmetry. Because of $w_{\mathbf{p}}^{\mu }$ is an essential part of the
argument of the Airy functions, the behavior of the wave functions is
significantly affected.

Fig. 3 shows the near parabolic dispersion relations of the two coupled
levels and their corresponding spin down and up sublevels, for the electric
and magnetic fields under consideration. It can be seen the $p_{y}$
paraboloids symmetry according to Eqs. (12) and (13). Although there is a
little difference between spin paraboloids, due to the large energy
difference between the resonant levels of both wells ($\sim $12 meV) and the
splitting of the spin sublevels ($\sim $0.01 meV at $p_{y}=0$) it is not
possible to distinguish minima behavior. Thus, we have enlarged in Fig. 4
the bottom of the pair of paraboloids (spin down and spin up) for the ground
level. As expected, both paraboloids shift in opposite $p_{x}$ directions
resulting in a sublevel anticrossing. This displacement is due not only to
the magnetic field but also to the electric one (12).

Considering a $p_{y}=0$ section of the former figure we can get a more
accurate 2D representation (Fig. 5) of the anticrossing, minima $p_{x}$
position, and energy splitting. The inset displays anticrossing area
enlargement. In this kind of anticrossing the slope of $\varepsilon _{%
\mathbf{p}\sigma }^{\mu }$ varies without changing the sign. However, as we
will see below for the low magnetic fields under study, we work in the
region where Van Hove singularities remain in the density of states. In
order to have a more detailed overview of the anticrossing region we have
also included in Fig. 6 the contour plot around anticrossing for different
constant energy values.

Next, we analyze the density of states $\rho \left( \varepsilon \right) $.
This function is related to several spin and interwell tunneling properties.
Also, it is proportional to the photoluminescence excitation (PLE)
intensity, one of the most used techniques to get information of quantum
structures \cite{23}. As mentioned before, we have found remains of the Van
Hove singularities for fields under consideration. So, we have used magnetic
field intensities varying from $0$ to $0.1$ T to analyze singularities
behavior. The shape of $\rho \left( \varepsilon \right) $ is shown in Fig.
7. Note that, when $H=0$ T, energy paraboloids shift a certain amount $\mp 
\mathrm{w}_{\mathbf{p}}^{\mu }=\pm \bar{\mathrm{v}}^{\mu }p$ because of the $%
\bar{\mathrm{v}}^{\mu }$ dependence on the electric field. In this case,
because both paraboloids bottom are at the same energy, we have clear $%
\varepsilon ^{-1/2}$ type singularities in each subband. As expected, these
peaks disappear gradually by growing magnetic field because of the different
vertical paraboloids shift, which leads to a greater slope at the
anticrossing point. In turn, interfaces contribute with a manifested delay
in the quenching of the $\rho \left( \varepsilon \right) $ singularities, as
well as an additional broadening of these peaks. That is, although the
singularities should only appear at zero magnetic field, they still remain
at the band anticrossing position for low magnetic fields, as shown in Fig.
7. Another significant feature is that the $\rho \left( \varepsilon \right) $
anticrossing peak is softened when increasing barrier height, disappearing
for lower magnetic fields than used in this work\cite{3}

Finally, Fig. 8 shows PLE spectra for different Gaussian halfwidth $\gamma $
at a fixed $H=0.1$ T. For the magnetic field used before ($H=0.01$ T) the
two adjacent peaks, corresponding to each resonant pair of states, overlap.
Thus, we have used a magnetic field ten times higher because this field
allows us to tell the peaks apart for small $\gamma $ values. Evolution of
the first two peaks with $\gamma $ is depicted in Fig. 9. As can be seen,
PLE peaks corresponding to the two different spin transitions are still
distinguishable for $\gamma $\ values {}{}beyond $1$ meV. These results show
the same general behavior than available experimental data\cite{24}.

\section*{4. Conclusions}

In this paper we have analyzed the electron spin behavior in narrow-gap
ACQWs under transverse electric and in-plane magnetic fields, including the
role of abrupt interfaces. We used the Kane model with nonsymmetric boundary
conditions, caused by the two different wells width, to solve the eigenvalue
problem. Based in this model and the transfer matrix approach we have
performed a useful tool to tackle any layered structure with abrupt
interfaces and subjected to different perturbations. To do that we have
implemented $8\times 8$ matrices for the boundary conditions to describe
near-parabolic dispersion relations. The model allows us the study of the
spin peculiarities of levels anticrossing.

Because interface contributions oppose intrinsic spin-orbit effect,
mechanisms that mix the Pauli contribution with the two kinds of spin-orbit
contributions (from a low magnetic field and from heterojunctions) are
different. As a result, numerical calculations for $InGaAs-InAlAs$
structures lead to magnetoinduced variations of the energy dispersion
relations, under very low in-plane magnetic fields. This effect is
particularly \ appreciable at anticrossings. As dispersion relations are
used to obtain the density of states, this function is also affected by the
abrupt interfaces: a new kind of $\rho \left( \varepsilon \right) $\
singularity, which does not disappear when increasing magnetic fields,
appears. Furthermore, there is an energy broadening of the peaks.

Finally, we have calculated PLE intensity because, as mentioned before,
mid-infrared PLE spectroscopy is a suitable technique to find
characteristics of energy spectrum\cite{23}. This technique provides direct
information of the energy spectrum when interband transitions are modified
by in-plane magnetic fields.

Since, for our purposes, a thorough analysis of the band structure little
contributes to describe eigenstates, the main conclusions of this work
remain valid. We have also used along the work the assumption that the
confined potential is well described. Thus, the effective transverse field $%
F_{\bot }$ provides correct estimations both for magnetoinduced changes and
for the character of the dispersion laws. However, more detailed numerical
calculations are needed to describe the kinetic behavior. We will return to
this point in a forthcoming work where we will analyze the spin dynamics in
ACQWs. In summary, we have shown the importance of the contribution from the
interfaces, in narrow gap structures, to the spin polarization through the
electronic dispersion relations, and how this contribution can be detected
by PLE. The model can be extended to the analysis of the negative
magnetoresistance\cite{25} as well as peculiarities of the spin current
through ADQW\cite{26}

\newpage \appendix{}

\section{Boundary conditions}

After a first integration of the Schr\"{o}dinger equation over an
heterojunction at $z=L$, we get the third class boundary conditions:%
\begin{equation}
\left. \frac{\widehat{p}_{z}}{m_{w}}\Psi ^{w}(\mathbf{p,}z)\right\vert
_{z=L}-\left. \frac{\widehat{p}_{z}}{m_{b}}\Psi ^{b}(\mathbf{p,}%
z)\right\vert _{z=L}-i\chi \left[ \widehat{\mathbf{\sigma }}\times \mathbf{p}%
\right] _{z}\left. \Psi ^{w}(\mathbf{p,}z)\right\vert _{z=L}=0,  \label{A1}
\end{equation}%
together with the wave function continuity at interfaces%
\begin{equation}
\left. \Psi ^{w}(\mathbf{p,}z)\right\vert _{z=L}-\left. \Psi ^{b}(\mathbf{p,}%
z)\right\vert _{z=L}=0.  \label{A2}
\end{equation}%
which lead to a set of four equations we can write as%
\begin{eqnarray}
&&\left. \frac{-i}{m_{w}}\frac{\partial }{\partial z}\Psi ^{w}(\mathbf{p,}%
z)\right\vert _{z=L}+\left. \frac{i}{m_{b}}\frac{\partial }{\partial z}\Psi
^{b}(\mathbf{p,}z)\right\vert _{z=L}+\frac{\chi }{\hbar }\left[ 
\begin{array}{cc}
0 & p_{-} \\ 
-p_{+} & 0%
\end{array}%
\right] \left. \Psi ^{w}(\mathbf{p,}z)\right\vert _{z=L}=0,  \notag \\
&&\left. \Psi ^{w}(\mathbf{p,}z)\right\vert _{z=L}-\left. \Psi ^{b}(\mathbf{%
p,}z)\right\vert _{z=L}=0.  \label{A3}
\end{eqnarray}%
After substitution of $\Psi ^{\mu }(\mathbf{p,}z)$ by $\psi ^{\mu \sigma }(%
\mathbf{p})\varphi ^{\mu \sigma }(z)$ in (A3) we obtain:%
\begin{eqnarray*}
&&\left. \left\{ a_{b+}Ai(\xi _{\mathbf{p}+}^{b})+b_{b+}Bi(\xi _{\mathbf{p}%
+}^{b})+\rho _{\mathbf{p}-}^{b}\left[ a_{b-}Ai(\xi _{\mathbf{p}%
-}^{b})+b_{b-}Bi(\xi _{\mathbf{p}-}^{b})\right] \right\} \right\vert _{z=L}
\\
&=&\left. \left\{ a_{w+}Ai(\xi _{\mathbf{p}+}^{w})+b_{w+}Bi(\xi _{\mathbf{p}%
+}^{w})+\rho _{\mathbf{p}-}^{w}\left[ a_{w-}Ai(\xi _{\mathbf{p}%
-}^{w})+b_{w-}Bi(\xi _{\mathbf{p}-}^{w})\right] \right\} \right\vert _{z=L},
\end{eqnarray*}%
\begin{eqnarray*}
&&\left. \left\{ \rho _{\mathbf{p}+}^{b}\left[ a_{b+}Ai(\xi _{\mathbf{p}%
+}^{b})+b_{b+}Bi(\xi _{\mathbf{p}+}^{b})\right] +a_{b-}Ai(\xi _{\mathbf{p}%
-}^{b})+b_{b-}Bi(\xi _{\mathbf{p}-}^{b})\right\} \right\vert _{z=L} \\
&=&\left. \left\{ \rho _{\mathbf{p}+}^{w}\left[ a_{w+}Ai(\xi _{\mathbf{p}%
+}^{w})+b_{w+}Bi(\xi _{\mathbf{p}+}^{w})\right] +a_{w-}Ai(\xi _{\mathbf{p}%
-}^{w})+b_{w-}Bi(\xi _{\mathbf{p}-}^{w})\right\} \right\vert _{z=L},
\end{eqnarray*}%
\begin{eqnarray*}
&&\left. \left( \frac{m_{e}}{m_{b}}\left\{ a_{b+}Ai^{\prime }(\xi _{\mathbf{p%
}+}^{b})+b_{b+}Bi^{\prime }(\xi _{\mathbf{p}+}^{b})+\rho _{\mathbf{p}-}^{b}%
\left[ a_{b-}Ai^{\prime }(\xi _{\mathbf{p}-}^{b})+b_{b-}Bi^{\prime }(\xi _{%
\mathbf{p}-}^{b})\right] \right\} \right) \right\vert _{z=L} \\
&=&\left( \frac{m_{e}}{m_{w}}\left\{ a_{w+}Ai^{\prime }(\xi _{\mathbf{p}%
+}^{w})+b_{w+}Bi^{\prime }(\xi _{\mathbf{p}+}^{w})+\rho _{\mathbf{p}-}^{w}%
\left[ a_{w-}Ai^{\prime }(\xi _{\mathbf{p}-}^{w})+b_{w-}Bi^{\prime }(\xi _{%
\mathbf{p}-}^{w})\right] \right\} \right. \\
&&\left. +\left. \frac{i\chi }{\hbar }\left\{ \rho _{\mathbf{p}+}^{w}p_{-}%
\left[ a_{w+}Ai(\xi _{\mathbf{p}+}^{w})+b_{w+}Bi(\xi _{\mathbf{p}+}^{w})%
\right] +p_{-}\left[ a_{w+}Ai(\xi _{\mathbf{p}+}^{w})+b_{w+}Bi(\xi _{\mathbf{%
p}+}^{w})\right] \right\} \right) \right\vert _{z=L},
\end{eqnarray*}%
\noindent 
\begin{eqnarray}  \label{A4}
&&\left. \left( \frac{m_{e}}{m_{b}}\left\{ \rho _{\mathbf{p}+}^{b}\left[
a_{b+}Ai^{\prime }(\xi _{\mathbf{p}+}^{b})+b_{b+}Bi^{\prime }(\xi _{\mathbf{p%
}+}^{b})\right] +a_{b-}Ai^{\prime }(\xi _{\mathbf{p}-}^{b})+b_{b-}Bi^{\prime
}(\xi _{\mathbf{p}-}^{b})\right\} \right) \right\vert _{z=L}  \notag \\
&=&\left( \frac{m_{e}}{m_{w}}\left\{ \rho _{\mathbf{p}+}^{w}\left[
a_{w+}Ai^{\prime }(\xi _{\mathbf{p}+}^{w})+b_{w+}Bi^{\prime }(\xi _{\mathbf{p%
}+}^{w})\right] +a_{w-}Ai^{\prime }(\xi _{\mathbf{p}-}^{w})+b_{w-}Bi^{\prime
}(\xi _{\mathbf{p}-}^{w})\right\} \right.  \notag \\
&&-\left. \left. \frac{i\chi }{\hbar }\left\{ p_{+}\left[ a_{w+}Ai(\xi _{%
\mathbf{p}+}^{w})+b_{w+}Bi(\xi _{\mathbf{p}+}^{w})\right] +\rho _{\mathbf{p}%
-}^{w}p_{+}\left[ a_{w+}Ai(\xi _{\mathbf{p}+}^{w})+b_{w+}Bi(\xi _{\mathbf{p}%
+}^{w})\right] \right\} \right) \right\vert _{z=L}.  \notag \\
\end{eqnarray}%
The matrix form of the above equations for coefficients $a_{\mu \sigma }$, $%
b_{\mu \sigma }$ leads to expression (20). 

\newpage 

\begin{figure}[tbp]
\begin{center}
\includegraphics{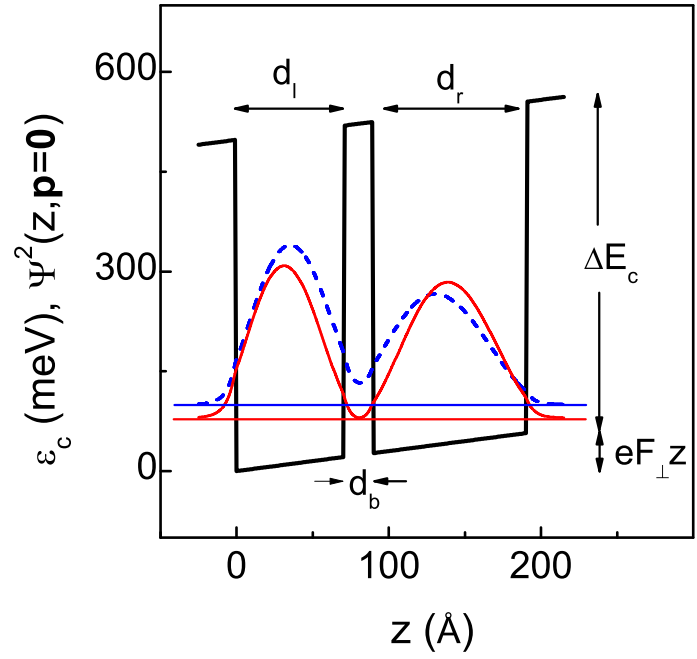}
\end{center}
\par
\addvspace{-1 cm}
\caption{(Online color) Conduction band diagram for undoped asymmetric
double quantum well. Horizontal thin lines show electron energy levels and
thin curves correspond to squared wave functions close to the resonance.}
\end{figure}
\newpage 

\begin{figure}[tbp]
\begin{center}
\includegraphics{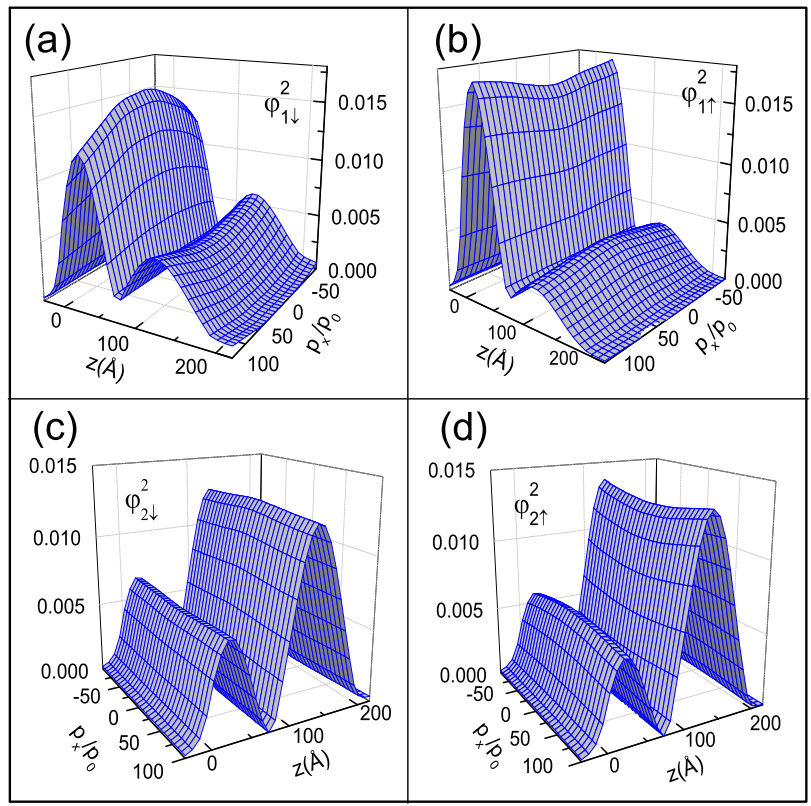}
\end{center}
\par
\addvspace{-2cm}
\caption{(Online color) Squared wave functions vs normalized $p_{x}/p_{0}$
momentum for the first spin resonant levels. }
\end{figure}
\newpage

\begin{figure}[tbp]
\begin{center}
\includegraphics{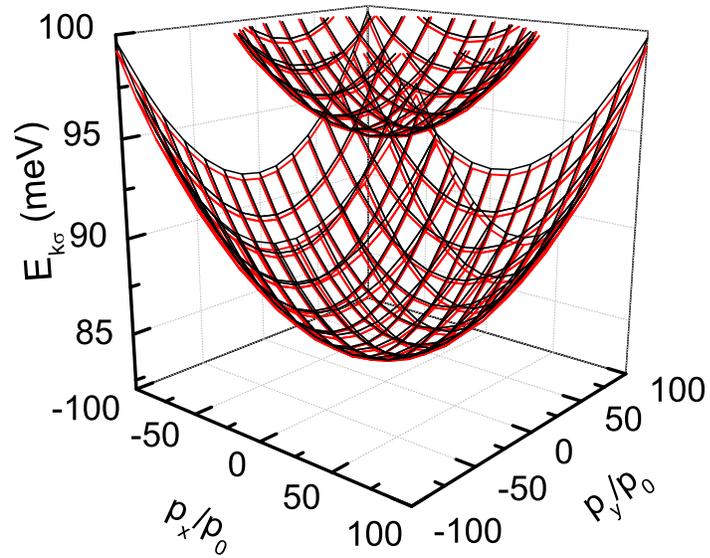}
\end{center}
\par
\addvspace{-3 cm}
\caption{(Online color). Near parabolic dispersion relations of the two
coupled levels close to the resonance, and their corresponding spin down
(black) and spin up sublevels. }
\end{figure}
\newpage

\begin{figure}[tbp]
\begin{center}
\includegraphics{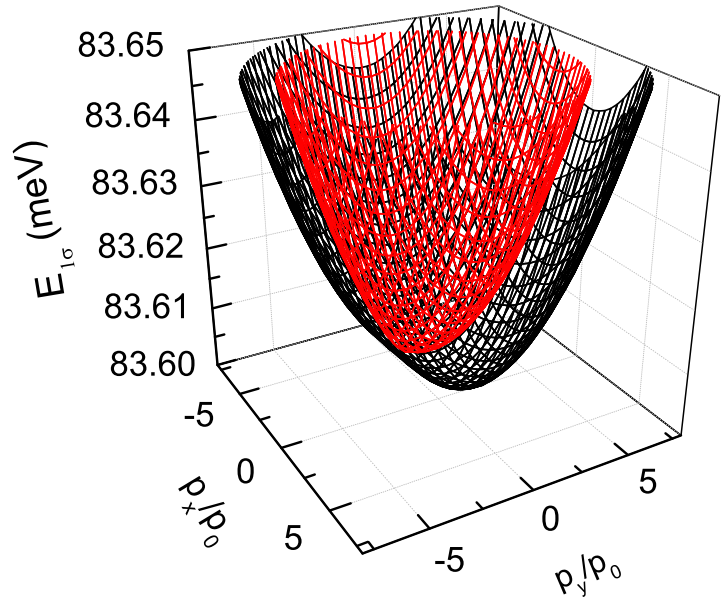}
\end{center}
\par
\addvspace{-3 cm}
\caption{(Online color) Bottom of the two deepest paraboloids just after
resonance and detail of spin down (black) and spin up anticrossing. Levels
are mainly located in the narrow (left) QW. }
\end{figure}
\newpage

\begin{figure}[tbp]
\begin{center}
\includegraphics{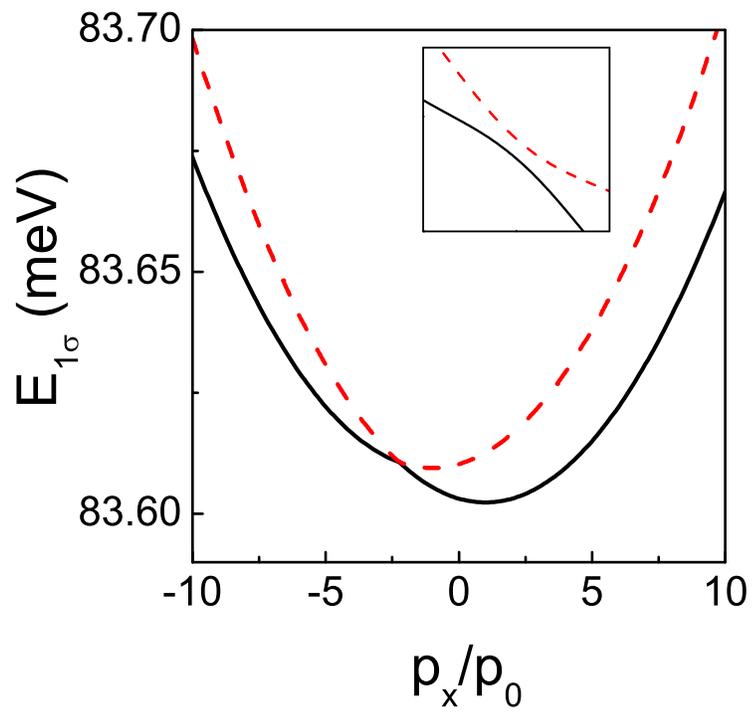}
\end{center}
\par
\addvspace{-1 cm}
\caption{(Online color) Two dimensional dispersion relations for $%
p_{y}/p_{0}=0.$ Inset shows a magnified image of the spin down and spin up
anticrossing. Solid line; spin down; dashed line; spin up.}
\end{figure}
\newpage

\begin{figure}[tbp]
\begin{center}
\includegraphics{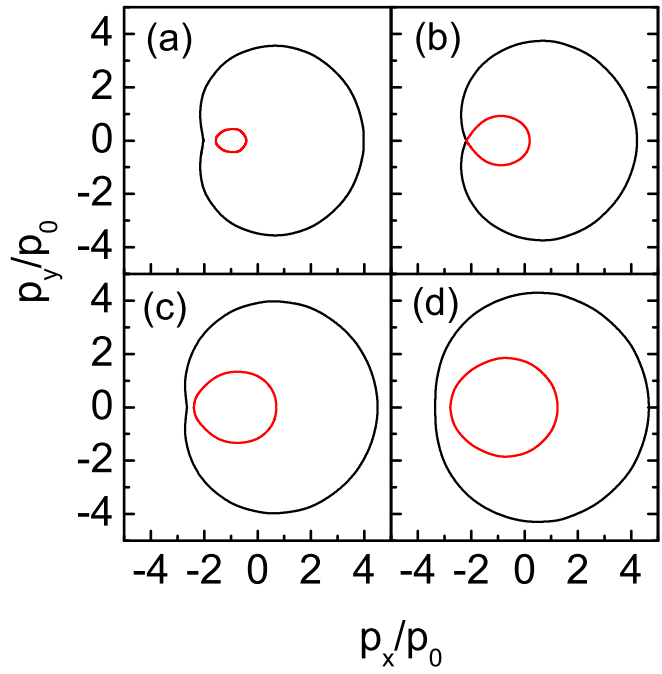}
\end{center}
\par
\addvspace{-1 cm}
\caption{(Online color). Contour plot of the deepest energy levels around
anticrossing. (a) $E_{1\protect\sigma }=83.610$ meV; (b) $E_{1\protect\sigma %
}=83.611$ meV; (c) $E_{1\protect\sigma }=83.612$ meV; and (d) $E_{1\protect%
\sigma }=83.614$ meV. Outer line: spin down; inner line: spin up.}
\end{figure}
\newpage

\begin{figure}[tbp]
\begin{center}
\includegraphics{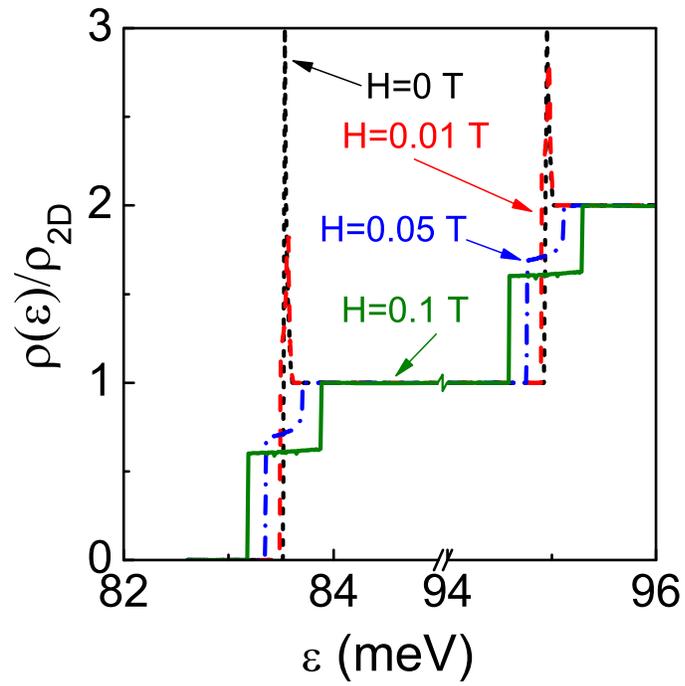}
\end{center}
\par
\addvspace{-1 cm}
\caption{(Online color) Density of states for the same electric field of
previous figures and different magnetic fields.}
\end{figure}
\newpage

\begin{figure}[h]
\begin{center}
\includegraphics{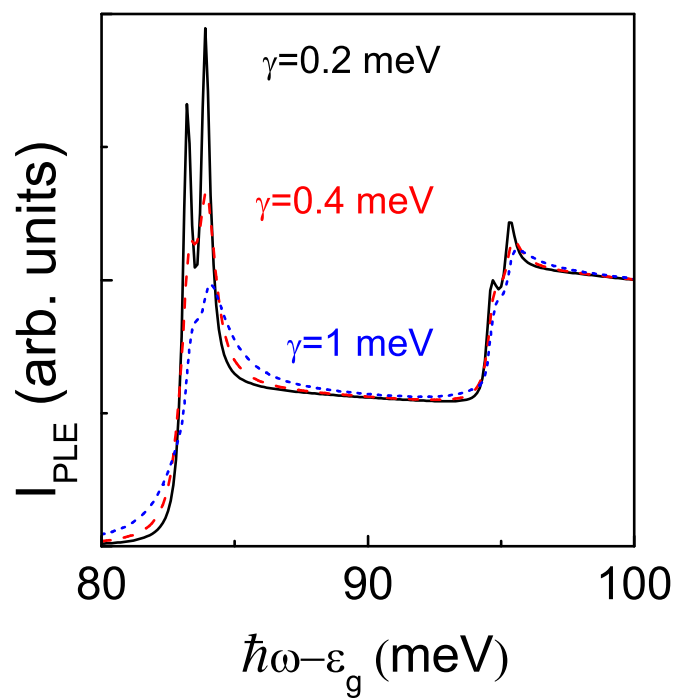}
\end{center}
\par
\addvspace{-1 cm}
\caption{(Online color) Photoluminescence excitation intensity for $H=0.1$ T
and different $\protect\gamma $ values.}
\end{figure}
\newpage

\begin{figure}[h]
\begin{center}
\includegraphics{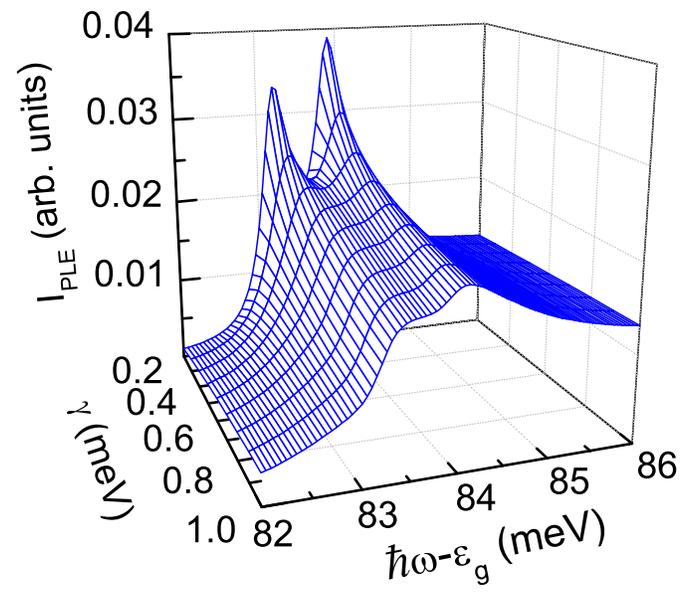}
\end{center}
\par
\addvspace{-1 cm}
\caption{(Online color) Photoluminescence excitation vs $\protect\gamma $
corresponding to the two first transitions. Parameters are the same as in
Fig. 8.}
\end{figure}

\end{document}